\pdfoutput=1
\documentclass[preprint2,numberedappendix,iop]{emulateapj-rtx4}

\usepackage{graphicx,graphics,amsmath}
\usepackage{natbib}
\usepackage{bm,url}
\usepackage{times}
\usepackage{amssymb}
\usepackage{color}
\usepackage{float}
\usepackage{arydshln}
\usepackage[export]{adjustbox}
\usepackage{csvsimple}
\graphicspath{./png/}
\usepackage{color}
\usepackage{multirow}
\def\blue{\textcolor{black}}

\newcommand{\Fig}[1]{Figure~\ref{#1}}
\newcommand{\Figs}[2]{Figures~\ref{#1} and \ref{#2}}
\newcommand{\Tabs}[2]{Tables~\ref{#1} and \ref{#2}}
\newcommand{\Eq}[1]{Equation~(\ref{#1})}

\newcommand{\Tab}[1]{Table~\ref{#1}}

\newcommand{\Eqs}[2]{equations~(\ref{#1}) and~(\ref{#2})}
\newcommand{\App}[1]{Appendix~\ref{#1}}

\newcommand{\we}{Waldmeier Effect}
\newcommand{\II}{{\sc \romannumeral 2}}

\shorttitle{WALDMEIER EFFECT IN STELLAR CYCLES}
\shortauthors{GARG ET AL.}

\begin{document}

\title{Waldmeier Effect in Stellar Cycles}

\author{Suyog Garg}
\affil{Indian Institute of Information Technology, Design and Manufacturing, Kancheepuram, Chennai 600127, India}

\author{Bidya Binay Karak} 
\affil{Department of Physics, Indian Institute of Technology (Banaras Hindu University), Varanasi 221005, India}

\author{Ricky Egeland}
\affil{High Altitude Observatory, National Center for Atmospheric Research, 3080 Center Green Dr., Boulder, CO 80301, USA}

\author{Willie Soon}
\affil{Harvard-Smithsonian Center for Astrophysics, Cambridge, MA 02138, USA}

\and
\author{Sallie Baliunas}
\affil{Retired, Harvard-Smithsonian Center for Astrophysics, Cambridge, MA 02138, USA}

\begin{abstract}
One of the most robust features of the solar magnetic cycle is that the stronger cycles rise faster than the weaker ones. This is popularly known as the Waldmeier Effect, which is known for more than 80 years. This fundamental feature of the solar cycle has not only practical implications,
e,g., in predicting the solar cycle, but also implications in understanding the solar dynamo.
Here we ask the question whether the Waldmeier Effect exists in other Sun-like stars.
To answer this question, we analyze the Ca \II{} H \& K S-index from Mount Wilson Observatory for 21 Sun-like 
\blue{
G--K stars.
}
We specifically check two aspects of Waldmeier Effect, namely, WE1: the anti-correlation between the rise times and the peaks
and WE2: the positive correlation between rise rates and amplitudes.
We show that except HD~16160, HD~81809, HD~155886 and HD~161239, all stars considered in the analysis show WE2.
While WE1 is found to be present only in some of the stars studied. Further, the WE1 correlation is weaker than the WE2. Both WE1 and WE2 exist in the solar S-index as well.  
Similar to the solar cycles, the magnetic cycles
of many stars are asymmetric about their maxima. 
The existence of the Waldmeier Effect and asymmetric cycles in Sun-like stars suggests that the dynamo mechanism which operates in the Sun is also operating in other stars.
\end{abstract}



\section{Introduction}
\label{sec:int}
The solar magnetic activity increases and decreases cyclically with an average
period of 11 years, however, both the duration and amplitude vary cycle to cycle. There are also
short-term variations within the cycle \citep{LB89}, making it difficult
to predict the future activity.
Nonetheless, there are some special features of the solar cycle.
One of these is that the rate of increase of the activity is not the same
for all cycles, rather it depends on the strength of the cycle.
This leads to a famous relation, called the Waldmeier Effect \citep{Wald}.
It says that the stronger cycles rise
faster (and take less time to reach their peak activities) than the weaker ones,
and vice-versa.
The \we\ has been extensively studied in different proxies of solar activity data
since its discovery in 1935.
However, limitations in different data sets sometimes make it difficult to establish
its existence. For example, \citet{Dik08} did not find a significant anti-correlation 
between rise times and amplitudes and claimed that the \we\ is not present
in the sunspot area data.
Later, a more careful analysis by \citet{KC11} computed the rise time and showed that \we\
exists in all, sunspot area, number and 10.7-cm radio flux data.
They further showed that this anti-correlation between rise times and amplitudes is one aspect of \we\ which they referred to as WE1.
The other aspect is the positive correlation between the rise rates and
the amplitudes and they referred to it as WE2.
It turned out that the latter feature is much more robust
and probably more useful \citep{CS08}. For example, when the \we\ is used to predict
the future solar cycle, WE2 is meant rather than WE1
which requires rise time of the cycle \citep[e.g.,][]{CS07,Kane08,RL12}.
Only after the cycle has reached its maximum can the rise time be obtained
while the rise rate can easily be obtained shortly after the cycle has started.

Many other stars, so-called the Sun-like stars (with spectral types F--M), which have convection zones in their outer layers produce magnetic fields through dynamo action 
\blue{
\citep{Pa55,Moffatt78,Gil83,O03,C14,Noyes84a,SBZ93,SB99}. 
}
Many of these stars indeed show clear magnetic cycles with varying amplitudes and durations \citep{Baliu95}. 
 Therefore, the important question that we want to answer in this paper is whether the \we\ can be seen in other stars as well. Checking the \we\ in other stars, however, is challenging because typically
we do not have a long dataset of magnetic activity. Among
the available observables of magnetic activity, Ca \II{} H \& K line emission (hereafter simply HK) is the longest
dataset that has been used to study the long-term variation of stellar activity.
The magnetic (non-thermal) heating in the chromosphere causes emission in the cores of 
HK. This HK emission is used to measure the stellar magnetic activity due to its strong correlation with the magnetic flux, as realized in the Sun \citep{Le59,Skumanich75,Sch92}.

The HK Project of Mount Wilson Observatory (MWO) regularly monitored more than 100 Sun-like
stars with spectral types early F to M, starting from 1966. However, the
Project ended in 2003. 
Therefore, this limited data of 37 years may be used to analyze the features of stellar cycles. Analyzing these data from 21 stars for which good cycles are seen, we shall explore the existence of \we\ in different Sun-like stars for the first time. Other features of stellar cycles, 
particularly the asymmetry of the cycles will also be identified. To put the Sun in this context, we shall also present the
analysis of \we\ using the same Ca \II{} HK proxy obtained from Kodaikanal Observatory, which was not done in the past.


\begin{figure*}
\includegraphics[scale=0.33]{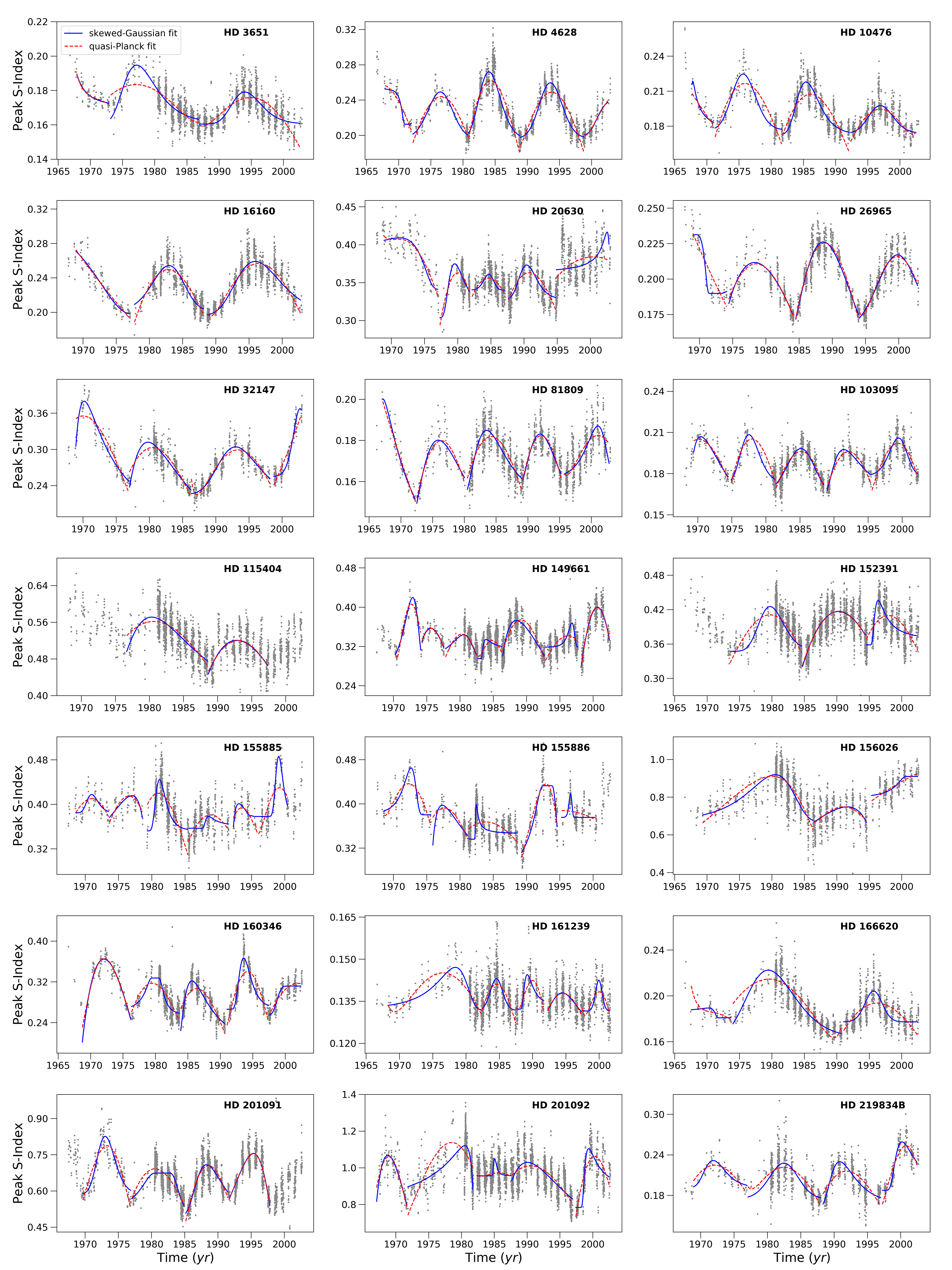} 
\caption{Stellar cycles used in our analysis. Dotted points 
are the MWO's calibrated S-index, while solid blue and dashed red lines are
the fitted data using skewed-Gaussian and quasi-Planck functions, respectively.}
\label{fig:panel_stars}
\end{figure*}

\begin{figure}
\includegraphics[scale=0.30]{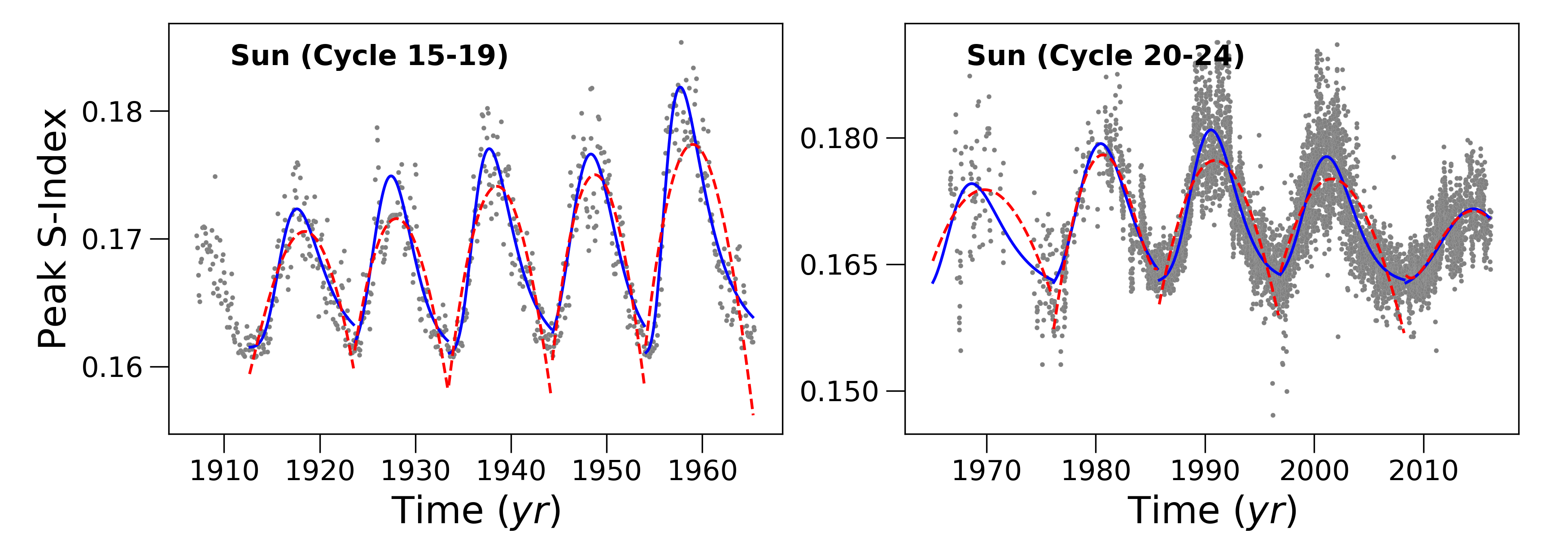} 
\caption{Same as \Fig{fig:panel_stars} but for the Sun.
Left panel: Kodaikanal Observatory data and right: MWO data scaled; see \citet{Ege17} for details.
}
\label{fig:panel_sun}
\end{figure}


\label{sec:data}


\section{Data Analyses}
\label{sec:analysis}
We use the fully calibrated S-index time series from the MWO HK Project.  
\blue{
In our analysis we need high-quality data having at least two consecutive unambiguous cycles.
These are available only in G and K stars which are also listed in \citet{BV07} and \citet{Sch13}.
From these lists, we find that only 21 stars are useful in our study. 
Chromospheric variations of these stars are shown
in \Fig{fig:panel_stars}, while the cycle properties are listed in \Tab{tab:summary}.
}
The dotted points in this figure show the raw data, the calibrated S-index.
The S-index for the Sun, obtained from the Kodaikanal Observatory is also shown for comparison in \Fig{fig:panel_sun}.

To study the \we\ using these S-index data, we need to compute the rise time, rise rate,
and peak value of each cycle. However, as can be seen from \Fig{fig:panel_stars}, the data are irregular (due to the rotation, active region emergence and decay, and uneven observations). Thus, we end up getting huge uncertainties in the cycle parameters if they are computed directly from the calibrated S-index. Therefore,
instead of using the raw calibrated S-index data, we produce approximate cycles of the data using the given profile(s)
and then the best-fitted cycles are used to study the \we.
What profile we use to produce the fitted data is guided by the fitting of solar data.
A popular fitting profile of the solar cycle is the following quasi-Planck
function \citep{HWR94}.

\begin{equation}
f(t) = \frac{a(t-t_0)^3}{\exp\left[\frac{(t-t_0)^2}{b^2}\right]-c},
\label{eq:planck}
\end{equation}

where $t$ is the time in year, $t_{\rm 0}$ is the starting time of the cycle, 
parameters $a$, $b$ and $c$ are related to the cycle amplitude (peak S-index),
the rise time, and the cycle asymmetry, respectively.
On the other hand, \citet{Du11} has demonstrated that the following skewed-Gaussian function
captures the solar cycle profile slightly better.

\begin{equation}
f(t) = A \exp \left[ \frac{-(t-t_m)^2}{2B^2[1+\alpha(t-t_m)]^2} \right] + f_{\mathrm {min}},
\label{eq:gauss}
\end{equation}
where $A$ is related to the cycle amplitude,
$t_{\rm m}$ is approximately the time of maximum,
$B$ relates to the width of the cycle rising phase,
and $\alpha$ is the asymmetry parameter.
The last term $f_{\rm min}$, which measures the offset S-index at the cycle minimum, 
was not present in the
\citet{Du11} model, but was introduced recently by \citet{Ege17}.
Although they preferred this profile, they have shown the fitting with the quasi-Planck function produces comparable results.
We, therefore, fit the stellar S-index with both profiles and study the \we\
separately in each fitted data. 

As the fitting will be performed for each cycle separately, we need to isolate the cycles first.
To do so, we bin the raw data with one-year intervals and then smooth with a three-point running window.
In this smoothed S-index, the cycles are isolated by considering the data in between two successive minima.
The individual cycle periods computed from two successive minima and their mean periods for all stars 
are enlisted in the last two columns of \Tab{tab:summary}.
Taking the minimum time as a heuristic value, we proceed to fit the raw S-index of individual cycles
with functions given in \Eqs{eq:planck}{eq:gauss}.
For the fitting, we use the Python \texttt{scipy} library \texttt{curve\_fit} routine.
This built-in function uses a Trusted Region Reflective (TRR)
method with the Levenberg–Marquardt (LM) algorithm applied
to trusted-region subproblems \citep[][p.\ 105]{more78}.
In \Figs{fig:panel_stars}{fig:panel_sun}, the solid and dashed lines show the fitted profiles using skewed-Gaussian (\Eq{eq:gauss})
and quasi-Planck function (\Eq{eq:planck}), respectively.
To measure a goodness of fit we compute the reduced $\chi^2_{\rm red}$. 
In evaluating $\chi^2_{\rm red}$, we consider the error as $1.2\%$ of the S-index as suggested in \citet{Baliu95}. 
\Tabs{tab:fitParams1}{tab:fitParams2} in \App{App}
list out the values of various fitted parameters and $\chi^2_{\rm red}$ for each of the stars used in the analysis. 
While for most stars these $\chi^2_{\rm red}$ values are reasonable considering that the models do not address short-timescale variability present in the data, 
we get significantly large values for a few stars where even the long-term variability appears to be not well represented by the models.
While for some stars these $\chi^2_{\rm red}$ values
are reasonable, we get significantly large values for a few stars which are due to poor data quality.
We further observe that the $c$ column for quasi-Planck fit contains some extreme negative values. 
As such, although the trend is seen in general for most of the stellar cycles, 
it cannot be said that the parameter $c$ corresponds to the cycle asymmetry. 
It can also be seen that the parameter $t_0$ does not increase monotonically in 
every consecutive cycle of a star. 
Thus, the correspondence between $t_0$ and the cycle start time is also not thorough.

Sadly, not all the cycles are usable for the analysis. 
The observational data of most of the stars contains a partial cycle
either initially or at the end of the observation period. These partial cycles cannot be used. Further, some cycles of HD 155885, HD 155886  and HD 201092 are unusable because of ill-fitting or inconsistent variability. \Tab{tab:fitParams2} of \App{App} lists out the parameters values for such unusable cycles, whereas \Tab{tab:fitParams1} gives the values for the usable cycles. The plots in \Fig{fig:panel_stars} can be reproduced from \Eqs{eq:planck}{eq:gauss} using the parameters given in \Tab{tab:fitParams1} and \Tab{tab:fitParams2}.
To compare the goodness of the fitting of different stars we further compute $\chi^2_{\rm red}$ 
for all the usable cycles in each star. This total $\chi^2_{\rm red}$  is shown in \Tab{tab:summary}.
For the solar cycles, we get $\chi^2_{\rm red}$ values as 3.5 for quasi-Planck and 2.5 for skewed-Gaussian fittings. 
%

We note that the observational data of the stellar magnetic activity 
sometimes contains local minima between the cycles. 
We ignore the local minima when the time difference between the minimum and the preceding maximum is much less compared to such difference for other cycles.
The threshold value of the time difference for this criterion of ignoring the local minima varies from star to star and depends on the cycle characteristics.
For instance, for HD 201092, the local minima are ignored if the time interval containing the minimum is $\sim3$ years or less. Whereas, in the  case of HD 149661, since the actual cycle durations are in that range, we do the same for a much smaller interval ($\sim1$ years).

Further, another important parameter to look into is the cycle asymmetry. Solar magnetic 
cycles are usually found to be asymmetric about the maxima. Therefore it is interesting to find out 
whether the stellar cycles are also asymmetric. 
A measure of this asymmetry in the cycle can be obtained by calculating the skewness of the data
defined as

\begin{equation}
\gamma_1 = \frac{ \sum_{i=1}^{N}(f^d_i-\Bar{f})^3 } { (N-1) \, \sigma_{fit}^3 }, 
\label{eq:skew}
\end{equation}

where $\Bar{f}$ is the mean strength of the cycle and $\sigma_{\rm fit}$ is the standard deviation of the cycle data from the fitted profile. 
Column 7 of \Tab{tab:summary} tabulates the mean value of the skewness for the stars $\overline{\gamma_1}$, calculated using the individual skewness values of usable cycles. 
A positive (negative) skewness means that the part of the cycle on the right (left) of the cycle maxima is skewed compared to the other side. 
For reference, the mean skewness of the solar cycles computed from our S-index is $0.3373$.

\begin{table*}
\caption{Summary of data analysis and results. }
\begin{center}
\resizebox{\textwidth}{!}{
\begin{tabular}{lcccccrlllc}
\hline
Star & SpecType & Data duration & \# cycles &  $\chi^2_{\rm red,P}$~~&~~$\chi^2_{\rm red,G}$ & $\overline{\gamma_1}~~$ & WE1? & WE2?& $P_{\rm cyc}\rm{ (yr)}$ & $P_{\rm cyc}^{\rm mean} \rm{(yr)}$ \\[2pt]
\hline \hline
HD 3651   & K0V   & 1966--2002 & 2 & 12.6 & 11.0 & 2.26 & WE1$^{\rm G}$ & WE2$^{\rm G}$  & 13.81,  15.99 & 14.90 \\
HD 4628   & K2V   & 1966--2002 & 3 & 20.5 & 11.6 & 0.60 & WE1$^{\rm {P,G}}$ & WE2$^{\rm {P,G}}$ &  8.67,  8.08, 9.98 & 8.91 \\
HD 10476  & K1V   & 1966--2002 & 3 & 16.5 & 10.7 & 0.85 & No & WE2$^{\rm {P,G}}$  & 10.50, 10.57, 10.61 & 10.56 \\
HD 16160  & K3V   & 1967--2002 & 2 & 18.6 & 17.8 & 0.00 & WE1$^{\rm G}$ & No  &  10.99, 14.31  & 12.65 \\
HD 20630  & G5V   & 1967--2002 & 3 & 11.0 & ~9.5 & 0.10 & WE1$^{\rm G}$ & WE2$^{\rm {P,G}}$ & 4.83, 5.50, 7.21 & 5.85 \\
HD 26965  & K1V   & 1967--2002 & 3 & ~~8.9 & ~8.8 & 0.02 & No & WE2$^{\rm {P,G}}$ &  10.08,  9.58, 8.96 & 9.54 \\
HD 32147  & K3+V  & 1967--2002 & 2 & 21.2 & 19.2 & 0.08 & WE1$^{\rm {P,G}}$ & WE2$^{\rm {P,G}}$ &  9.33, 12.42 & 10.88 \\
HD ${81809}^{\dag}$ & G2IV+G2V 
                  & 1966--2002 & 4 & ~~9.8 & ~8.4 & 0.14 & No & No & 7.96, 8.74, 6.50, 7.32 & 7.64 \\
HD 103095 & KV    & 1968--2002 & 4 & ~~9.1 & ~8.2 & 0.50 & No & WE2$^{\rm {P,G}}$ &  6.58, 7.50, 6.58, 6.90 & 6.89 \\
HD 115404 & K2.5V & 1968--2002 & 2 & 25.8 & 25.4 & $-0.19$& WE1$^{\rm {P,G}}$ & WE2$^{\rm G}$ & 11.99, 8.84
& 10.42 \\
HD 149661 & K0V   & 1967--2002 & 7 & 23.0 & 20.6 & 0.34 & No & WE2$^{\rm {P,G}}$  & 3.95, 3.54, 4.84, 3.75, 6.42, 5.34, 4.46
& 4.61 \\
HD 152391 & K8+V  & 1966--2002 & 3 & 28.7 & 24.6 & $-0.15$ & WE1$^{\rm G}$ & WE2$^{\rm {P,G}}$  & 11.16, 9.83, 8.10  & 9.70 \\
HD 155885 & K1V   & 1967--2002 & 4 & 29.3 & 19.7 & 0.30 & No & WE2$^{\rm P}$  & 5.08, 5.67, 6.19, 1.47  & 4.60 \\
HD 155886 & K2V   & 1967--2002 & 2 & 25.4 & 23.7 & 0.30 & WE1$^{\rm P}$ & No & 7.33, 5.99  & 6.67 \\
HD 156026 & K5V   & 1966--2002 & 2 & 75.3 & 74.1 & $-0.58$ & No & WE2$^{\rm P}$ & 17.16, 8.87  & 13.02 \\
HD 160346 & K2.5V & 1966--2002 & 4 & 28.2 & 16.4 & 0.22 & No & WE2$^{\rm {P,G}}$  & 7.58, 7.58, 7.13, 6.45 & 7.19 \\
HD 161239 & G2IIIb& 1966--2001 & 5 & ~~8.0 & ~7.6 & 0.55 & No & No & 14.17, 5.17, 4.67, 5.17, 4.29 & 6.69 \\
HD 166620 & K2V   & 1966--2002 & 2 & 22.4 & 17.4 & 5.99 & WE1$^{\rm {P,G}}$ & WE2$^{\rm G}$  & 17.08, 11.59 &14.34 \\
HD 201091 & K5V   & 1967--2002 & 4 & 27.7 & 28.5 & 0.39 & No & WE2$^{\rm {P,G}}$  & 7.42, 8.08, 6.50, 6.52  & 7.13 \\
HD 201092 & K7V   & 1967--2002 & 3 & 56.9 & 39.6 &$-0.07$& No & WE2$^{\rm {P}}$  & 4.76, 10.58, 5.32 & 6.89 \\
HD 219834B& K2V   & 1967--2002 & 4 & 50.5 & 38.0 & 0.36 & WE1$^{\rm G}$ & WE2$^{\rm {P,G}}$ &  8.10, 11.25, 8.92, 5.47 & 8.45 \\
Sun       & G2V   & 1913--2016 &10 & ~~3.5 &  ~2.5 & 0.34 & WE1$^{\rm {P,G}}$ & WE2$^{\rm {P,G}}$ & 7.84--11.49 & 10.35  \\
\hline\hline
\end{tabular}
}
\end{center}
\tablecomments{Here, `\# cycles' is the total number of usable cycles present in the data duration. $\chi^2_{\rm red,P}$ and $\chi^2_{\rm red,G}$ are the reduced $\chi^2$ computed over all usable cycles of a star fitted with 
quasi-Planck (\Eq{eq:planck}) and skewed-Gaussian (\Eq{eq:gauss}) profiles, respectively. $\overline{\gamma_1}$ is the mean Skewness $\gamma_1$ (a measure of the cycle asymmetry) for the usable cycles. Further, the superscripts on the WE1 and WE2 tell the fitting in which the correlation is found. Here, P stands for quasi-Planck fit, while G stands for skewed-Gaussian. The spectral type information is taken from \citet{ol16}. 
$^\dag$HD 81809 is an unresolved binary and the observed activity cycles are most likely linked to G2 subgiant \citep{Ege18}.
}
\label{tab:summary}
\end{table*}


\section{Results}
\label{sec:res}

\subsection{Checking the Waldmeier Effect in solar S-index}
\label{sec:sun}
As referred in the Introduction that previous authors have studied
the \we\ using sunspot number and area data, but not using any chromospheric magnetic proxy.
Therefore, we shall first check the \we\ for the Sun with the same type of chromospheric magnetic proxy that we are using for the other stars. 
This will show the robustness of our analysis and also put the Sun
in the stellar context.
For this, we use the S-index of Sun from MWO data for the last 5 cycles and 
Kodaikanal data where the data is available for cycles 15--19. 
As described in \citet{Ege17}, these two data sets are made homogeneous
by doing the following scaling.

\begin{figure*}
\centering
\includegraphics[scale=0.6]{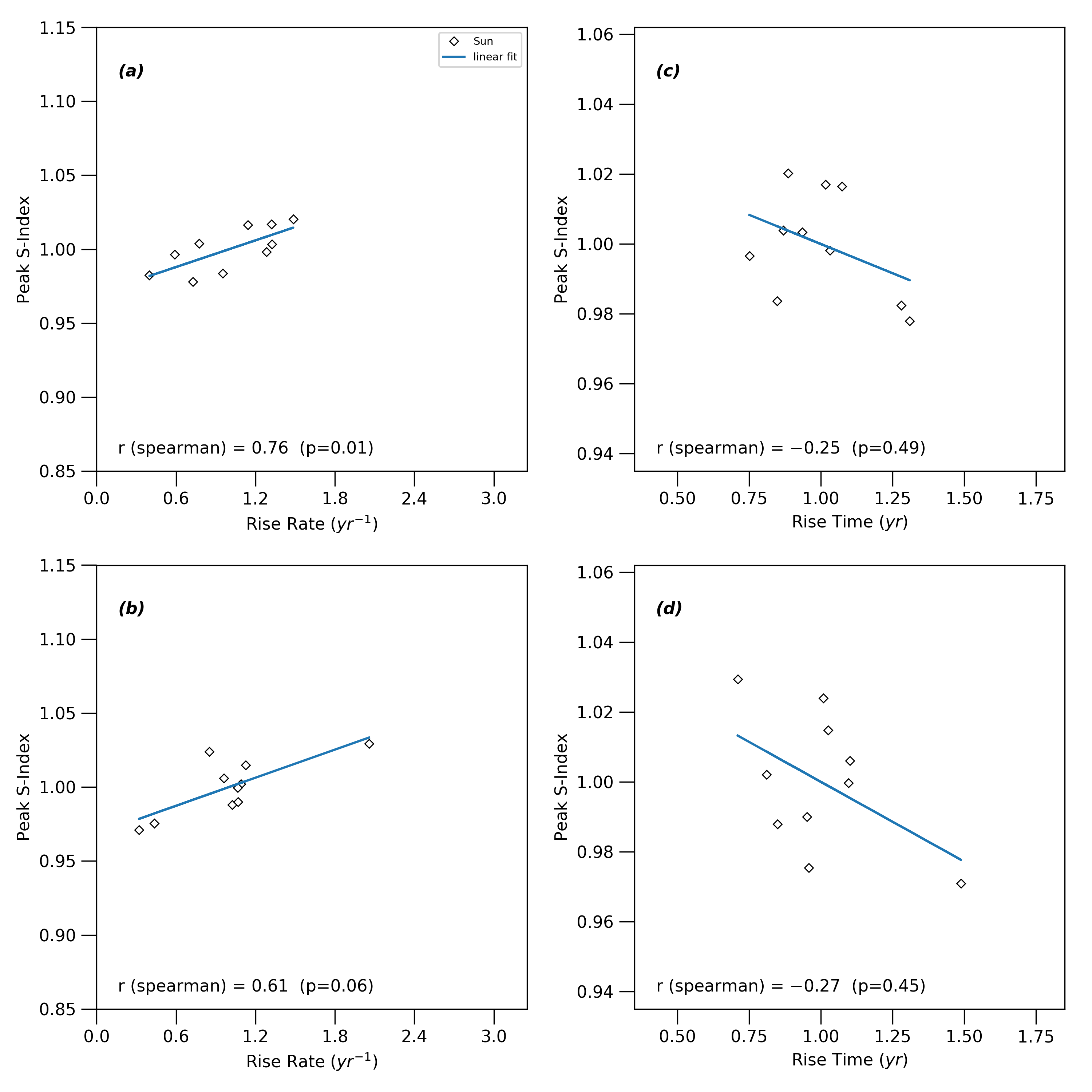}
\caption{
Left: scatter plots for WE2 (positive correlation between rise rates and the amplitudes).
Right: WE1 (anti-correlation between rise times and amplitudes) for the solar Cycles.
Top and bottom panels obtained from quasi-Planck and skewed-Gaussian fitted data, respectively.
\blue{
The solid lines are linear fits with slopes 
$m = 0.030 \pm 0.023$, $0.032 \pm 0.022$, $-0.033 \pm 0.062$ \& $-0.045 \pm 0.064$ and 
intercepts $c = 0.970 \pm 0.025$, $0.968 \pm 0.024$ , $1.033 \pm 0.063$ \& $1.046 \pm 0.065$,
with the rms-deviation of fittings being 
$0.011$, $0.013$, $0.014$ \& $0.018$, respectively for panels a, b, c and d.
}
}
\label{fig:sun}
\end{figure*}

The Kodaikanal data is in Integrated Sunlight Spectrometer (ISS) flux scale, calibrated by \citet{Be16} 
using the synoptic Ca \II{} {\it K} plage index from spectroheliograms from the Kodaikanal (KKL) Observatory in India. This time series data of Ca \II{} {\it K} emission is first transformed to the Ca \II{} {\it H} \& {\it K} measurement scale of Sacramento Peak National Observatory (NSO/SP) flux scale by applying following equation.
\begin{equation}
    K_{\rm KKL(SP)}=1.1388K_{\rm KKL(ISS)}-0.0071.
    \label{eq:trans1}
\end{equation}
Now, another transformation is performed between the NSO/SP {\it K} emission index and the MWO S-index. Assuming a linear relationship of the form, $S(K)=a+bK$, and using the cycle 23 fit parameters for the Sun (\citet{Ege17}), the following linear transformation is obtained:
\begin{equation}
   S(K)=(1.5\pm0.13)K + (0.031\pm0.013) 
   \label{eq:trans3}
\end{equation}
Using \Eq{eq:trans3} for the Kodaikanal data between cycles 15--19, homogenizes the solar data.

Previous studies have shown that the
WE2 (anti-correlation between rise rates and amplitudes) is a robust feature of the solar cycle.
Therefore, we analyze the WE2 first using the data obtained from the quasi-Planck fitting.
Following \citet{KC11}, we define the rise rate of a cycle as the change in the activity in a single year
and calculating the change between two consecutive years after leaving out one year
from the cycle beginning (minimum).
The scatter plot between the rise rate and the peak S-index is shown in the left panel of \Fig{fig:sun} both for the quasi-Planck and skewed-Gaussian fittings.
We find that the correlation is good and the value of the correlation coefficient
is comparable to the previous study \citep{CS08,KC11}.

Now, we consider WE1 which is the anti-correlation between the rise time and the strength.
As argued in \citet{Dik08} and \cite{KC11}, incorrect identification of cycle minimum
or maximum can lead to a large error in the rise time computed. We, therefore, consider the rise time
as the time taken by a cycle to grow from $20\%$ to $80\%$ of its peak.
The scatter plot between the rise time and the peak S-index is shown in the right panel of 
\Fig{fig:sun}.
As found in the other proxies of the solar cycle, the WE1 correlation is much weaker.
The correlation coefficients found in our S-index data is comparable to the previous 
values obtained from other proxies of the solar data \citep{Dik08, KC11}.


\begin{figure}
\centering
\includegraphics[scale=0.35]{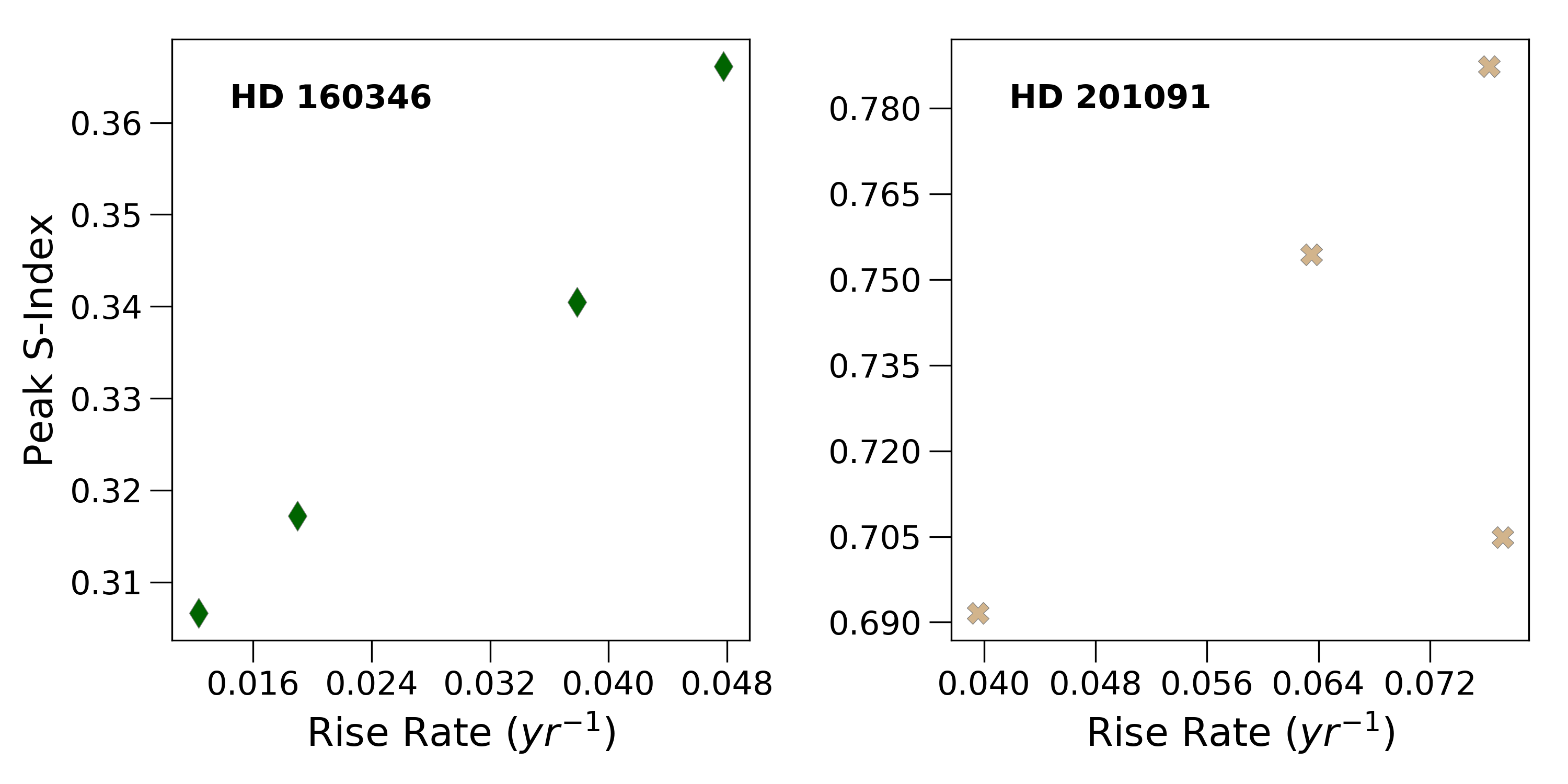}
\caption{WE2 obtained from the quasi-Planck fitted S-index data of stars HD 160346 (left) and HD 201091 (right).
}
\label{fig:we2example}
\end{figure}

\begin{figure}
\centering
\includegraphics[scale=0.6]{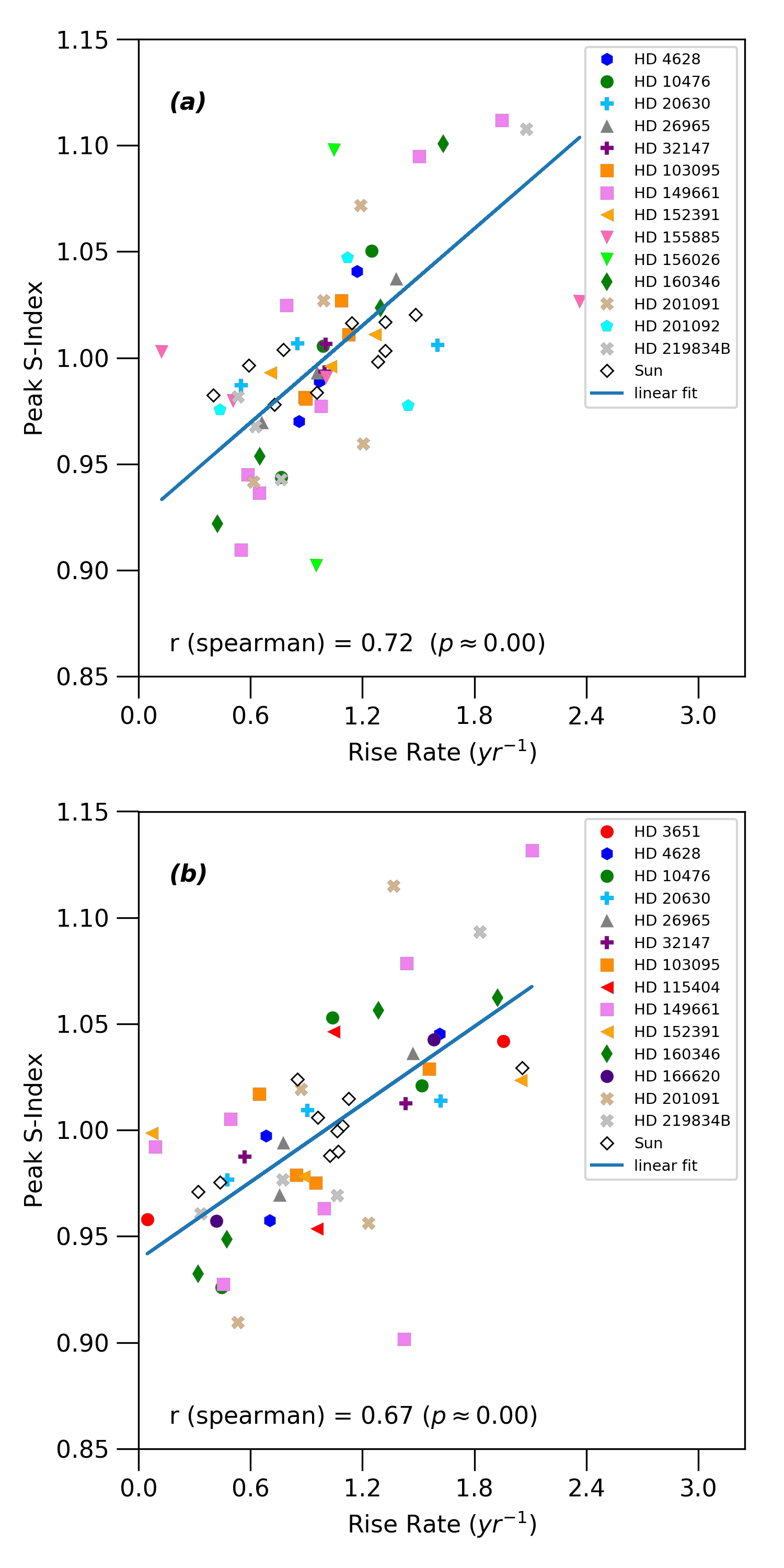}
\caption{
Combined scatter plot for WE2. 
Different symbols represent different stars. The linear Spearman correlation and confidence level
are printed on each plot.
Top and bottom panels obtained from quasi-Planck and skewed-Gaussian fitted data.
We note that the data of stars HD~16160, HD 81809, HD 155886, and HD 161239 are not included because these stars do not
show a positive correlation. 
\blue{
The solid lines are best linear fits with slopes $m = 0.076 \pm 0.025$ \& $0.061 \pm 0.021$ and intercepts $c = 0.924 \pm 0.028$ \& $0.939 \pm 0.024$, 
with the rms-deviation of fittings being $0.038$ \& $0.039$, respectively for top and bottom panels.
}
}
\label{fig:we2}
\end{figure}

\subsection{Checking the Waldmeier Effect in Stars}
\label{sec:wald}

Now we study the \we\ for the stars in the same way as we have done for the Sun. 
We first show the results of WE2 
 using the data obtained from the quasi-Planck fitting.
For the representation, the scatter plots between the rise rates and 
the peak S-indexes for HD 160346 and HD 201091 are shown in \Fig{fig:we2example}.
This shows that the stronger cycles rise faster than the weaker ones---establishing
the WE2 in these two stars. As the number of cycles in each star is very limited, the correlation coefficient
computed from the cycles of an individual star is not very meaningful. Therefore, we combine the data for 
all the stars which 
individually show the correlation for WE2. 
For a star,
if the Spearman correlation coefficient between rise rates and amplitudes is at least $0.1$,
then we mark that star to `hold WE2'.
This information is listed in \Tab{tab:summary}. 
\Tab{tab:WE_table} in \App{App} details the peak S--index and rise rate values for all the stars used in the analysis.
Surprisingly, we find that all the stars except HD~16160, HD~81809, HD~155886 and HD~161239, 
follow WE2 for at least one type of fitting. 

For HD~16160, there are only two cycles (see \Fig{fig:panel_stars}).
Therefore, with this limited data, it is difficult to make a conclusive comment about whether HD 16160 truly does not follow WE2.
The same argument may hold for HD~155886 because this star also has only two usable cycles
and both cycles are very noisy. In HD~161239, the correlation is destroyed due to the first cycle which is very different than the rest. As seen in \Fig{fig:panel_stars},
for HD~161239, the first cycle is the strongest one and the activity level during the minimum is relatively high. This produces an unexpectedly slow rise rate in the first cycle
and breaks the WE2 relation. If we do not consider the first cycle, then the star beautifully show WE2
(The linear Spearman correlation is $0.80$ for quasi-Planck fitted data 
and $0.20$ for skewed-Gaussian fitting).
In the case of HD 81809, the cycles are very regular and the data quality is good. 
However, it is very surprising that the star does not follow WE2.
The individual Spearman correlation coefficient between the rise rates and the amplitudes is $-0.80$ for the quasi-Planck fitting and $-0.40$ for the skewed-Gaussian. Thus the correlation is opposite to the one we expect for the WE2 to hold.

The combined data of stars having a good positive correlation between the rise rate and
the amplitude are shown in the top panel of \Fig{fig:we2}. 
We note that different stars have
different activity levels and cycle growth rates. Thus, to put all values in a comparable range, 
we normalize the data of individual stars with their average values. This does not affect the value of the 
correlation coefficient of the combined data.
For comparison, the data from the Sun is also shown in the same plot.
As seen from the top panel of \Fig{fig:we2}, the composite data show a good correlation between rise rates and strengths;
excluding the solar data, the linear Spearman correlation coefficient 
being $0.72$.
The data obtained from the skewed-Gaussian fitting show a similar behavior, 
although the correlation coefficient is lower; see bottom panel of \Fig{fig:we2}.
Interestingly, the skewed-Gaussian profile is fitting the data better than 
the quasi-Planck profile (less $\chi^2_{\rm red, G}$ values as seen from \Tab{tab:summary});
however, the Waldmeier relation is relatively weaker.

Finally, we explore WE1 for stars. 
Again, we first check the trend between rise times and the peaks for each star separately. 
We say that a star shows WE1 if the Spearman correlation coefficient between 
peaks and rise times is less than $-0.1$.
As listed in \Tab{tab:summary}, not all stars show the expected negative correlation.
We find that only five stars out of $21$ show WE1 for the quasi-Planck fitting. 
For the skewed-Gaussian fitting, four more stars show the solar-like WE1 correlation. 

\begin{figure}
\centering
\includegraphics[scale=0.6]{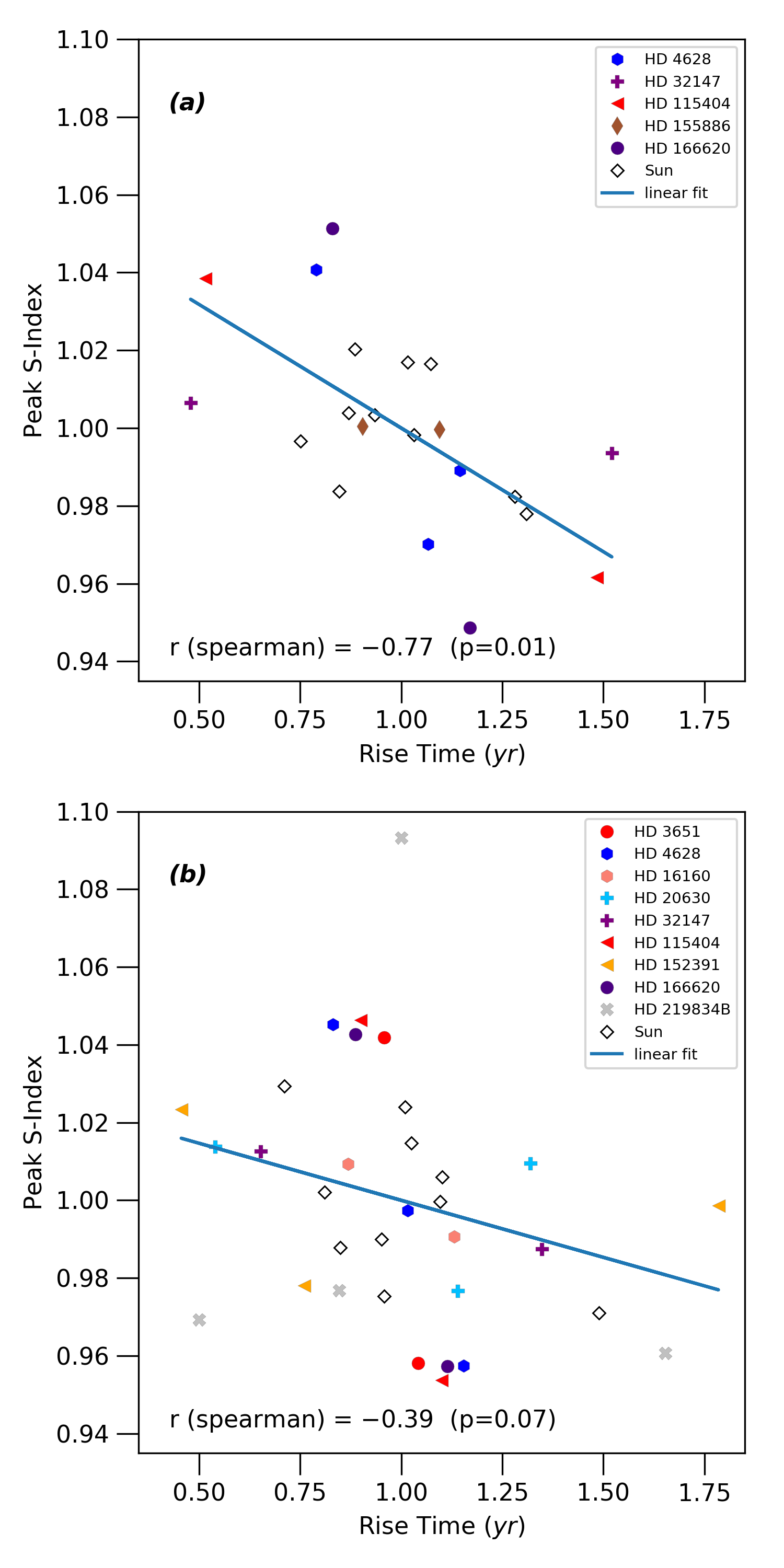}
\caption{
Scatter plot for WE1.
The figure format is same as \Fig{fig:we2}, however, in this case
number of stars following solar-like WE1 is less.
\blue{
The solid lines are linear fits with slopes $m = -0.064 \pm 0.055$ \& $-0.029 \pm  0.049$ and intercepts $c = 1.064 \pm 0.058$ \& $1.029 \pm 0.051$,  
with the rms-deviation of fittings being $0.026$ \& $0.036$, respectively for top and bottom panels.
}
}
\label{fig:we1}
\end{figure}

The scatter plots between rise times and peak S-index of all the stars
showing WE1 individually are shown in \Fig{fig:we1}. 
For comparison, the data for the Sun are overplotted in the same figure.
In this figure, we observe that the correlation coefficient is much weaker.
Excluding the solar data, the linear Spearman correlation coefficient is $-0.77$ ($p=0.01$)
for quasi-Planck fittings, while this value for skewed-Gaussian fitting is $-0.39$ ($p=0.07$). 
Note that in the latter case, the correlation is very less and thus the results are statistically insignificant.
It is the HD~219834B which has a very poor WE1 correlation ($-0.02$) and reduces the composite correlation
in \Fig{fig:we1}(b). If we exclude this star, then the correlation becomes $-0.52$ ($p=0.02$).
Weaker correlation for WE1 is not surprising because it is already known from the
study of solar data that WE1 is much weaker and it is sensitive to the data set in use
as well as the computation of rise times \citep{Dik08, KC11}.
Fortunately, if instead of defining rise time as the time 
taken by a cycle to grow from $20\%$ to $80\%$ of its peak, we define it using different values, then also we get
a solar-like negative correlation for WE1 from those stars which show correlations in \Fig{fig:we1};
(see \Tab{tab:WE1_corr}).
However, the correlation values for skewed-Gaussian remain far lower than those for the quasi-Planck fit.

\blue{
Let us discuss a consequence of WE2 relation. As shown in \Fig{fig:we2}, WE2 can be read as $S = m (\frac{dS}{dt}) + S_0$.
Now if we make a crude assumption: $\frac{dS}{dt} = \frac{ S}{ t}$, where $t$ is the total rise time,       
and if we also take the limit $\frac{m}{t} < 1$ (which is the case in our data), then to a first order,
we find: $S = (m S_0) \frac{1}{t} + S_0$. This relation fits the data
presented in \Fig{fig:we1} (WE1) poorly (rms deviation of the fittings being $0.029$ and $0.037$ for quasi-Planck and skewed-Gaussian fitted data). 
Our correlation plot for WE1 has already a large scatter and 
thus we cannot conclusively reject this variation over the linear variation 
represented as WE1.
}

\begin{table}[]
    \begin{center}
    \caption{WE1 correlation coefficients}
    \label{tab:WE1_corr}
    \begin{tabular}{cccccc}
    \hline \hline
    $t_{\rm start}$ & $t_{\rm end}$ & \multicolumn{2}{c}{$r_{\rm Spearman}$} \\
    \hline
    ~~~ & ~~~~ & quasi-Planck & skewed-Gaussian \\
    \hline
    0.15 & 0.75 & $-0.80$ & $-0.40$ \\
    0.15 & 0.80 & $-0.81$ & $-0.40$ \\
    0.15 & 0.85 & $-0.78$ & $-0.43$ \\
    0.20 & 0.75 & $-0.80$ & $-0.35$ \\
    0.20 & 0.80 & $-0.77$ & $-0.39$ \\
    0.20 & 0.85 & $-0.75$ & $-0.40$ \\
    0.25 & 0.75 & $-0.82$ & $-0.50$ \\
    0.25 & 0.80 & $-0.81$ & $-0.51$ \\
    0.25 & 0.85 & $-0.81$ & $-0.51$ \\
    \hline \hline
    \end{tabular}
    \end{center}
    \tablecomments{$t_{\rm end}$ and $t_{\rm start}$ are the times at which the activity is the given fraction of the peak value. The rise time is defined as the time difference between the $t_{\rm end}$ and $t_{\rm start}$.
    Stars used in the correlation computation are, for the quasi-Planck fitting case:
    HD~4628, HD~32147, HD~115404, HD~155886, and HD~166620, while for the skewed-Gaussian case: 
    HD~3651, HD~4628, HD~16160, HD~20630, HD~32147, HD~115404, 
    HD~152391, HD~166620, and HD~219834B.}
\end{table}


\section{Conclusion and Discussion}
\label{sec:conc}
In this study, we for the first time have explored the \we\ in Sun-like stars.
We have used MWO's calibrated S-index
\blue{
of $21$ stars having observations for at least two consecutive unambiguous cycles. These stars having
high-quality cycles belong to 
the spectral index of G--K (three G-stars and the rest are K-stars); 
see \citet{Egeland:2017:thesis} for the discussion on the cycle quality of MWO stars and \citet{Sch13} for an HR diagram of the cycling MWO stars.
}
After carrying out systematic
analyses of the data, we check two related aspects of the WE, namely, WE1 (the anti-correlation between
rise times and the amplitudes) and WE2 (the positive correlation between rise rates and amplitudes).
We find that all these stars, except HD~16160, HD~81809, HD~155886 and HD~161239,
show WE2 in at least one type of fittings used for the analysis.

For HD~16160, there are data for only two cycles with little variations in the cycle duration and amplitude
and the fitting of the data with the analytical profile is also poor. Thus we may expect that
when high-quality data are available, this star may show WE2 relation.
The same expectation can be held for HD~155886 because this star also has only two usable cycles
and both cycles are very noisy. Interestingly, in HD~161239  the first cycle is very different than the rest;
it is the strongest cycle and the activity level, even during the cycle minimum, is relatively high;
see \Fig{fig:panel_stars}. This produces an unexpectedly slow rise rate in the first cycle
and breaks the WE2 relation. If we do not consider the first cycle, then the star beautifully show WE2
(the linear Spearman correlation is $0.8$ for quasi-Planck fitted data and $0.2$ for skewed-Gaussian fitting).
HD~81809 is a peculiar star. In the available observation period, it has four excellent cycles
and the quality of the data
is fairly good as evident with good fittings with the analytical profiles (small $\chi^2_{\rm red}$).
Still, the star does not show WE. This possibly gives a hint that the periodicity of the
magnetic cycle in HD~81809 is caused by a different mechanism. 
Coincidently, HD 81809 is an unresolved binary system and the cyclic magnetic activity is possibly caused by the
A component which is a subgiant (with a rotation period of about 40 days and an effective temperature of 5757~K) \citep{Ege18}.
It could be that this slowly rotating subgiant possesses the so-called anti-solar differential rotation \citep{Har16}
and consequently, the dynamo mechanism could be fundamentally different than that in the other solar-like stars \citep{Kar15,KT19}.
Incidentally, the possibility of the existence of anti-solar differential rotation in slowly rotating stars
and the cyclic dynamos in those stars are highlighted in recent observational and theoretical works \citep{Kar15,viv19,W18,BG18,KT19}.
Clearly, further research is needed to identify the exact dynamo mechanism causing the excellent cycles in HD~81809
and why the \we\ is broken in this star.

In contrast to WE2, WE1 is observed to hold only in a small number of stars ($5$ out of $21$, for the quasi-Planck fitting).
For the Sun, we already know that the WE2 is a robust feature of the solar cycle and it exists in all
the proxies of the solar cycle, while WE1 correlation is very poor and sensitive to the data quality \citep{Dik08,KC11}.
Our analysis of the chromospheric activity of Sun also confirms that the WE1 correlation is weaker than the WE2.
Thus with our limited data sets for the Sun-like stars, and the existence of a strong correlation between
the rise rate and peak S-index (WE2), confirms that other Sun-like stars studied in this paper show
the \we.

The \we\ is a very distinctive and stringent feature of the solar cycle which has implication in predicting
the activity level of the following cycle \citep{CS07,Kane08,CS16, MKB17}.
Explaining this feature is a challenge to the solar dynamo models. However, a correct dynamo model
for the solar cycle must explain this special feature \citep{KC11,PK11,PC14,MKB17}.
Previously, using 25-year records of Ca \II{} HK emission fluxes of solar-type stars, \citet{SBZ94} showed
a similar type of inverse relationship between the amplitude of activity and the cycle length both for the
solar cycles and the ensemble of solar-type stars.
This result and the existence of \we\ in other Sun-like stars, as found in the present study,
suggest that the same mechanism is responsible for producing magnetic activity cycles in Sun and other Sun-like  
\blue{
G--K stars. 
}
Hence, the dynamo models which are successful in explaining the features of the solar cycle can be extended to other Sun-like stars.

It is also worth considering the converse implication; that is, if the
Sun and Sun-like stars operate under the same dynamo mechanism, then
observational constraints obtained for the stars are also applicable
to the Sun.  In particular, it has been observed from stellar data
that the cycle period increases along with the rotation
period \citep{SB99,BV07} along two branches separated by their
activity level and rotation period.  More recent re-analyses of the
MWO HK Project data have called into question whether such a
trend exists for the so-called ``active'' branch, stars with
$\log(R^\prime_{\rm HK}) \gtrsim -4.75$ and semi-empirical Rossby
number $\lesssim 1.5$ \citep{Olspert18,BoroSaikia18}.  The other,
``inactive'' branch, however, remains robust in these same studies,
and more importantly this cycle branch covers the solar
activity-Rossby parameter regime.  Both Babcock--Leighton
flux transport dynamos \citep{JBB10,KKC14} and recent global MHD
simulations \citep{Strugarek:2017, Strugarek:2018} are unable to
reproduce this observed cycle period trend 
\citep[however see][who find some correct trend in a pumping-dominated dynamo model]{Hazra19}.  
By necessity, our study
included only targets with high-quality cycles, which are the same set
of stars comprising the inactive branch.  By finding the WE operating
in these stars, it appears more likely that they are operating the
same dynamo mechanism as the Sun, and therefore viable models for the
solar cycle should produce an increasing cycle period with an
increasing rotation period.  
This indirect inference suggests a significant revision in the current dynamo models is required.


\begin{acknowledgements}
We thank Bibhuti Jha and Prasun Dutta for discussion on the data analysis.
We further thank the anonymous referee for careful review and providing comments.
SG thanks IIT (BHU) for providing one-month visit in the initial phase of the project.
BBK sincerely thanks SERB/DST for providing research grant through the Ramanujan Fellowship (project no SB/S2/RJN-017/2018).
RE was supported by the NCAR Advanced Study Program Postdoctoral Fellowship.  The National Center for Atmospheric Research is sponsored by the National Science Foundation.
\end{acknowledgements}

\bibliography{paper}


\appendix
\section{Supplementary material}
\label{App}


\begin{table*}[]
\centering
\caption{List of fit parameters for usable cycles of the stars}
\label{tab:fitParams1}
\resizebox{\textwidth}{!}{%
\begin{tabular}{lccccccrcccccr}
\hline \hline
\multirow{2}{*}{HD} & \multirow{2}{*}{Cycle \#} & \multirow{2}{*}{$t_{\rm start}$} & \multicolumn{5}{c}{quasi-Planck fit} & \multicolumn{6}{c}{skewed-Gaussian fit} \\
\cline{4-14}
 &  &  & $a$ & $b$ & $c$ & $t_0$ & $\chi^2_{\rm red}$ & $A$ & $B$ & $\alpha$ & $t_m$ & $f_{\rm min}$ & $\chi^2_{\rm red}$ \\[2pt]
 \hline
\multirow[t]{2}{*}{3651} & 1 & 1973.077 & -0.00003 & 21.96642 & 1.05552 & 1996.256 & 15.7 & 0.03397 & 2.52670 & 0.06877 & 1977.213 & 0.16074 & 14.3 \\
 & 2 & 1986.891 & 0.00005 & 18.42781 & 1.06091 & 1979.013 & 10.0 & 0.01864 & 2.26416 & 0.04595 & 1994.014 & 0.16042 & 8.1 \\
\multirow[t]{3}{*}{4628} & 1 & 1972.138 & 0.00037 & 13.55695 & -3.30787 & 1956.948 & 8.0 & 0.05232 & 1.88710 & -0.00608 & 1976.500 & 0.19686 & 6.7 \\
 & 2 & 1980.805 & 0.00068 & 9.06178 & 0.75710 & 1974.630 & 24.4 & 0.07622 & 1.64646 & 0.04085 & 1984.039 & 0.19581 & 15.2 \\
 & 3 & 1988.888 & 0.00049 & 9.46364 & 1.03405 & 1985.272 & 20.2 & 0.06302 & 1.76819 & 0.01515 & 1993.597 & 0.19657 & 9.3 \\
\multirow[t]{3}{*}{10476} & 1 & 1971.208 & 0.00017 & 13.61922 & 0.67148 & 1961.034 & 13.0 & 0.04810 & 1.84349 & 0.01893 & 1975.644 & 0.17655 & 9.4 \\
 & 2 & 1981.708 & 0.00015 & 14.02285 & 0.76646 & 1971.242 & 21.7 & 0.04520 & 1.71951 & 0.08420 & 1985.423 & 0.17267 & 10.9 \\
 & 3 & 1992.274 & -0.00019 & 11.91185 & 1.05735 & 2007.279 & 11.8 & 0.02371 & -1.88174 & 0.01890 & 1996.973 & 0.17389 & 11.0 \\
\multirow[t]{2}{*}{16160} & 1 & 1977.674 & 0.00019 & 13.93008 & 0.60479 & 1967.328 & 17.5 & 0.05286 & 2.36702 & -0.02248 & 1982.956 & 0.20142 & 16.1 \\
 & 2 & 1988.674 & 0.00019 & 13.16737 & 1.03675 & 1984.667 & 19.5 & 0.06283 & -3.39093 & 0.03947 & 1995.942 & 0.19624 & 19.1 \\
\multirow[t]{3}{*}{20630} & 1 & 1977.260 & 0.00136 & 7.94665 & 0.83796 & 1971.787 & 9.0 & 0.07114 & 1.20693 & 0.24111 & 1979.557 & 0.30363 & 8.4 \\
 & 2 & 1982.093 & 0.00036 & 12.33989 & 0.78578 & 1971.620 & 9.2 & 0.02087 & -0.73085 & -0.06690 & 1984.701 & 0.34015 & 8.2 \\
 & 3 & 1987.593 & 0.00031 & 13.20755 & 0.71826 & 1976.531 & 13.5 & 0.04495 & 1.28795 & 0.08658 & 1990.403 & 0.32820 & 11.2 \\
\multirow[t]{3}{*}{26965} & 1 & 1974.350 & 0.00011 & 15.62085 & 0.81033 & 1962.197 & 9.1 & 16.55227 & 74.08259 & 0.02908 & 1978.322 & -16.34062 & 9.1 \\
 & 2 & 1984.434 & 0.00026 & 11.74674 & 0.81468 & 1976.605 & 8.9 & 0.12171 & 4.35422 & 0.03059 & 1988.659 & 0.10445 & 8.4 \\
 & 3 & 1994.017 & 0.00032 & 10.38231 & 1.04316 & 1990.593 & 8.8 & 0.05403 & -3.01383 & -0.00346 & 1999.836 & 0.16298 & 9.0 \\
\multirow[t]{2}{*}{32147} & 1 & 1976.927 & 0.00029 & 12.59847 & 0.77504 & 1967.146 & 24.3 & 0.26280 & 4.66508 & 0.09866 & 1979.777 & 0.04906 & 21.5 \\
 & 2 & 1986.260 & 0.00036 & 11.13495 & 1.03585 & 1983.393 & 17.7 & 0.07866 & 2.95511 & 0.03546 & 1993.002 & 0.22544 & 16.6 \\
\multirow[t]{4}{*}{81809} & 1 & 1972.440 & 0.00016 & 12.66994 & 0.85374 & 1963.370 & 9.1 & 0.06014 & 3.73925 & 0.07183 & 1975.907 & 0.11997 & 8.7 \\
 & 2 & 1980.402 & 0.00011 & 14.85605 & 0.80285 & 1968.688 & 10.3 & 0.03604 & 2.45462 & 0.10103 & 1983.514 & 0.14879 & 7.7 \\
 & 3 & 1989.152 & 0.00025 & 11.15987 & 0.79085 & 1980.370 & 6.5 & 0.04123 & 2.50524 & 0.05027 & 1991.952 & 0.14185 & 6.3 \\
 & 4 & 1995.652 & 0.00018 & 11.83466 & 1.05833 & 1990.811 & 12.1 & 0.02616 & 1.48407 & -0.13072 & 2001.061 & 0.16082 & 11.1 \\
\multirow[t]{4}{*}{103095} & 1 & 1974.826 & 0.00037 & 10.19001 & 0.79966 & 1967.278 & 9.0 & 0.03542 & 1.36056 & 0.06468 & 1977.521 & 0.17296 & 7.7 \\
 & 2 & 1981.409 & -0.00021 & 12.63704 & 0.44887 & 1999.699 & 9.0 & 0.03368 & 2.03049 & -0.01620 & 1985.168 & 0.16457 & 8.5 \\
 & 3 & 1988.909 & 0.00026 & 11.15162 & 0.84375 & 1980.663 & 6.9 & 0.03869 & 2.20636 & 0.14569 & 1991.435 & 0.15881 & 5.9 \\
 & 4 & 1995.492 & -0.00030 & 10.93660 & 0.75811 & 2011.034 & 11.1 & 0.02731 & 1.28343 & -0.04186 & 1999.482 & 0.17863 & 9.5 \\
\multirow[t]{2}{*}{115404} & 1 & 1976.588 & 0.00010 & 22.58563 & 0.72117 & 1956.311 & 23.0 & 0.22537 & 5.22062 & 0.06002 & 1980.333 & 0.34573 & 22.4 \\
 & 2 & 1988.588 & 0.00028 & 14.96163 & 0.94085 & 1978.375 & 31.3 & 282.66608 & 205.69381 & 0.02673 & 1992.821 & -282.14557 & 31.2 \\
\multirow[t]{7}{*}{149661} & 1 & 1970.269 & 0.01221 & 3.81203 & 1.03289 & 1969.268 & 13.9 & 0.15751 & 1.04511 & -0.19775 & 1972.905 & 0.26208 & 8.1 \\
 & 2 & 1974.220 & 0.00229 & 6.62061 & 0.85257 & 1968.886 & 7.8 & 75.00468 & 52.97827 & 0.07636 & 1975.594 & -74.64756 & 7.9 \\
 & 3 & 1977.769 & -0.00169 & 7.01593 & 1.01273 & 1986.771 & 20.9 & 0.04939 & 1.47420 & -0.15638 & 1980.582 & 0.29442 & 20.7 \\
 & 4 & 1982.602 & 0.00396 & 5.14714 & 1.05998 & 1980.400 & 25.1 & 0.03914 & -0.75822 & 0.96105 & 1983.918 & 0.29514 & 23.7 \\
 & 5 & 1986.352 & 0.00111 & 8.54298 & 0.86251 & 1980.437 & 26.1 & 133.38530 & 83.86106 & 0.12729 & 1988.474 & -133.01263 & 22.5 \\
 & 6 & 1992.769 & -0.00056 & 10.54707 & 0.76645 & 2007.466 & 22.8 & 0.04965 & -0.44305 & -0.47344 & 1996.829 & 0.31828 & 17.8 \\
 & 7 & 1998.102 & 0.00584 & 4.88098 & 1.00684 & 1996.073 & 18.5 & 276.25825 & 81.97349 & -0.00568 & 2000.666 & -275.85840 & 18.4 \\
\multirow[t]{3}{*}{152391} & 1 & 1973.467 & -0.00019 & 17.06988 & 0.25778 & 1999.924 & 23.4 & 0.07811 & -1.91092 & 0.04456 & 1979.719 & 0.34722 & 22.6 \\
 & 2 & 1984.634 & 0.00024 & 14.61202 & 0.94103 & 1975.983 & 22.0 & 316.31972 & 243.45347 & 0.01403 & 1990.243 & -315.90312 & 21.9 \\
 & 3 & 1994.467 & 0.00030 & 13.80294 & 0.81915 & 1983.537 & 46.7 & 0.07751 & -0.81094 & 0.52461 & 1996.366 & 0.35840 & 31.4 \\
\multirow[t]{4}{*}{155885} & 1 & 1968.515 & -0.00049 & -11.68000 & 0.81894 & 1983.157 & 15.8 & 0.03299 & 0.83394 & 0.13242 & 1970.985 & 0.38463 & 14.7 \\
 & 2 & 1973.598 & -0.00075 & 10.12397 & 0.83278 & 1987.052 & 19.1 & 34.89182 & 45.08485 & -0.17104 & 1977.030 & -34.47625 & 18.4 \\
 & 3 & 1979.265 & 0.00084 & 10.03315 & 0.67185 & 1970.014 & 30.4 & 0.09207 & 0.73580 & 0.27387 & 1981.179 & 0.35220 & 20.1 \\
 & 6 & 1995.531 & 0.00290 & 6.25400 & 1.04317 & 1993.654 & 33.8 & 0.10820 & 0.54075 & 0.24093 & 1999.087 & 0.37812 & 20.4 \\
\multirow[t]{2}{*}{155886} & 1 & 1968.515 & 0.00146 & 7.88908 & 1.05315 & 1965.385 & 19.9 & 0.08489 & 0.86324 & -0.27408 & 1972.666 & 0.38021 & 13.6 \\
 & 4 & 1989.181 & 1.94477 & 7.53534 & -43424.72961 & 1970.264 & 28.4 & -0.20321 & -15.47080 & 0.16180 & 1999.423 & 0.43289 & 29.2 \\
\multirow[t]{2}{*}{156026} & 1 & 1969.237 & 2.65535 & 22.31880 & -1184502.3623 & 1902.567 & 60.4 & 0.25095 & 3.42172 & -0.05961 & 1980.496 & 0.66815 & 58.4 \\
 & 2 & 1986.404 & 0.00084 & 11.31426 & 1.05650 & 1981.170 & 108.5 & 261.25935 & 150.01094 & -0.07238 & 1991.608 & -260.51102 & 109.3 \\
\multirow[t]{4}{*}{160346} & 1 & 1968.657 & -0.02475 & -9.11479 & -425.93069 & 1992.820 & 15.9 & 596.91812 & 171.15102 & 0.05111 & 1972.061 & -596.55346 & 14.9 \\
 & 2 & 1976.240 & 0.00062 & 10.08616 & 0.70288 & 1968.566 & 31.9 & -0.07281 & 47.46115 & 0.10467 & 1989.547 & 0.32746 & 19.1 \\
 & 3 & 1983.823 & 0.00078 & 9.13790 & 0.74947 & 1976.597 & 27.3 & 0.16106 & 1.99316 & 0.21495 & 1985.622 & 0.16077 & 14.9 \\
 & 4 & 1990.957 & 0.00280 & 5.89130 & 1.02220 & 1988.825 & 25.4 & 0.10065 & 0.88474 & 0.23066 & 1993.598 & 0.26608 & 14.1 \\
\multirow[t]{5}{*}{161239} & 1 & 1968.207 & 0.00004 & 17.87804 & 1.06062 & 1961.118 & 6.9 & 0.01522 & 2.18507 & -0.11828 & 1978.408 & 0.13185 & 6.6 \\
 & 2 & 1982.374 & 0.00019 & 11.52722 & 0.54792 & 1971.630 & 10.0 & 0.01128 & -0.90487 & 0.00436 & 1984.579 & 0.13176 & 9.3 \\
 & 3 & 1987.540 & 0.00116 & 5.83240 & 1.06613 & 1985.029 & 7.0 & 0.01230 & 0.62130 & 0.53831 & 1989.209 & 0.13219 & 6.5 \\
 & 4 & 1992.207 & 0.00009 & 14.18839 & 0.80967 & 1979.752 & 6.1 & 0.01278 & 2.08953 & 0.04462 & 1994.462 & 0.12517 & 6.1 \\
 & 5 & 1997.374 & 0.00093 & 6.21154 & 1.06954 & 1994.774 & 8.2 & 0.01101 & 0.48922 & -0.01211 & 1999.959 & 0.13149 & 7.7 \\
\multirow[t]{2}{*}{166620} & 1 & 1973.907 & -0.00006 & -18.15521 & 1.03682 & 1995.824 & 23.2 & 0.05913 & -3.64608 & 0.02849 & 1979.446 & 0.16325 & 18.2 \\
 & 2 & 1990.990 & 0.00005 & 19.42901 & 0.78151 & 1975.706 & 21.3 & 0.02689 & -1.31451 & 0.00040 & 1995.684 & 0.17729 & 16.3 \\
\multirow[t]{4}{*}{201091} & 1 & 1969.587 & 0.01023 & 9.26666 & -34.92482 & 1956.381 & 32.9 & 0.23855 & 1.32241 & 0.05963 & 1973.026 & 0.58765 & 27.0 \\
 & 2 & 1977.003 & 0.00104 & 10.81097 & 0.81174 & 1969.221 & 26.9 & -0.22431 & 37.98379 & 0.07924 & 1994.379 & 0.67393 & 30.1 \\
 & 3 & 1985.087 & 0.00277 & 7.93770 & 0.73027 & 1980.048 & 30.8 & 0.21634 & 1.98884 & 0.09979 & 1988.300 & 0.49238 & 30.2 \\
 & 4 & 1991.587 & 3.32037 & 8.67855 & -70877.1084 & 1968.643 & 24.3 & 0.50460 & -2.93363 & -0.08084 & 1995.371 & 0.25069 & 24.1 \\
\multirow[t]{3}{*}{201092} & 1 & 1967.000 & 0.00889 & 6.10606 & 0.80022 & 1962.519 & 28.4 & 556.05306 & 70.15428 & 0.11064 & 1968.792 & -554.98578 & 28.1 \\
 & 2 & 1971.760 & -0.00118 & 12.23483 & 0.81287 & 1991.300 & 65.7 & 571.55755 & 76.11615 & -0.35400 & 1980.523 & -570.43734 & 43.3 \\
 & 5 & 1997.343 & 0.00655 & 6.72303 & 0.84075 & 1993.362 & 49.4 & 0.32077 & -0.95785 & 0.59137 & 1999.381 & 0.78356 & 36.2 \\
\multirow[t]{4}{*}{219834B} & 1 & 1968.849 & 0.00019 & 13.15917 & 0.75811 & 1958.099 & 16.8 & 0.03996 & 1.54044 & 0.09407 & 1971.896 & 0.19152 & 13.5 \\
 & 2 & 1977.015 & 0.00031 & 10.52077 & 1.05078 & 1973.266 & 52.3 & 0.05264 & 2.12248 & -0.00475 & 1982.494 & 0.17502 & 48.3 \\
 & 3 & 1988.265 & 0.00025 & 11.80385 & 0.79634 & 1979.495 & 64.1 & 0.06813 & 1.81362 & 0.15120 & 1990.869 & 0.16159 & 37.5 \\
 & 4 & 1997.182 & 0.00727 & 7.85046 & -58.37132 & 1985.728 & 24.8 & 0.07179 & 1.38579 & 0.24365 & 2000.202 & 0.18730 & 14.6 \\
 \hline \hline
\end{tabular}%
}
\end{table*}


\begin{table*}[]
\begin{center}
\caption{List of fit parameters for unusable cycles}
\label{tab:fitParams2}
\resizebox{\textwidth}{!}{%
\begin{tabular}{lccccccrcccccr}
\hline \hline
\multirow{2}{*}{HD} & \multirow{2}{*}{Cycle \#} & \multirow{2}{*}{$t_{\rm start}$} & \multicolumn{5}{c}{quasi-Planck fit} & \multicolumn{6}{c}{skewed-Gaussian fit} \\
\cline{4-14}
 &  &  & $a$ & $b$ & $c$ & $t_0$ & $\chi^2_{\rm red}$ & $A$ & $B$ & $\alpha$ & $t_m$ & $f_{\rm min}$ & $\chi^2_{\rm red}$ \\[2pt]
 \hline
3651 & 0 & 1967.808 & 0.00008 & 14.73875 & 1.11571 & 1960.119 & 4.1 & 0.03167 & 0.67457 & 0.90677 & 1967.530 & 0.15995 & 4.1 \\
\multirow[t]{2}{*}{4628} & 0 & 1967.805 & 0.00029 & 12.65300 & 0.28500 & 1952.529 & 10.5 & 0.03987 & 11.70434 & -0.24340 & 1967.233 & 0.21260 & 6.8 \\
 & 4 & 1998.870 & 0.00031 & 11.00919 & 1.04188 & 1994.962 & 11.2 & 0.03843 & 2.59672 & 0.10718 & 2003.165 & 0.20019 & 11.2 \\
10476 & 0 & 1967.624 & 184.34417 & 239.29236 & 220329.3196 & 696.276 & 7.2 & 34.03461 & 9.20649 & 2.06801 & 1967.805 & -33.81592 & 5.8 \\
16160 & 0 & 1968.841 & -2.08536 & -86.55002 & -3203538.7610 & 2043.742 & 11.7 & -0.10334 & -4.51996 & -0.02049 & 1977.208 & 0.30135 & 11.1 \\
\multirow[t]{2}{*}{20630} & 0 & 1968.843 & 0.00044 & 11.34389 & 1.09690 & 1962.617 & 14.9 & 0.06800 & 5.55820 & -0.12169 & 1971.215 & 0.34104 & 14.9 \\
 & 4 & 1994.800 & 0.00004 & 26.33570 & 0.89218 & 1973.889 & 34.7 & 0.06921 & 0.55279 & -1.00383 & 2002.515 & 0.34751 & 37.3 \\
26965 & 0 & 1968.934 & -0.00055 & 4305.49170 & -592.71992 & 2032.142 & 15.6 & 0.04155 & 2.26920 & -0.38731 & 1969.508 & 0.18979 & 5.9 \\
\multirow[t]{2}{*}{32147} & 0 & 1968.843 & 0.00033 & 13.10860 & 0.54002 & 1955.226 & 27.5 & 0.36062 & 2.64644 & 0.24537 & 1970.177 & 0.01928 & 12.6 \\
 & 3 & 1998.677 & 0.00785 & 3525.92019 & -824.44765 & 1969.406 & 19.2 & 0.11311 & 0.91264 & -0.08602 & 2002.682 & 0.25432 & 8.3 \\
81809 & 0 & 1967.068 & -0.24283 & 135.90863 & -225724.3452 & 2024.127 & 4.6 & 0.53697 & 3.96863 & 0.38283 & 1967.168 & -0.33681 & 4.4 \\
103095 & 0 & 1969.326 & 0.00016 & 13.69542 & 0.68319 & 1955.239 & 6.9 & 14.81514 & 33.93372 & 0.21716 & 1970.436 & -14.60824 & 6.2 \\
\multirow[t]{2}{*}{155885} & 4 & 1985.456 & 0.00041 & 11.99483 & 0.87560 & 1976.661 & 24.1 & 0.02134 & 0.67884 & 0.68844 & 1988.540 & 0.35696 & 34.2 \\
 & 5 & 1992.200 & 0.00166 & 7.68229 & 0.79647 & 1985.495 & 25.6 & 65.72575 & 36.34042 & 0.43686 & 1993.110 & -65.32469 & 24.7 \\
\multirow[t]{3}{*}{155886} & 2 & 1975.848 & 0.00056 & 11.12578 & 0.72436 & 1965.695 & 11.0 & 45.88451 & 39.58132 & 0.25605 & 1977.346 & -45.48709 & 10.7 \\
 & 3 & 1981.265 & 0.00012 & 18.00152 & 0.81553 & 1965.094 & 35.8 & 0.06358 & 0.17859 & 2.82572 & 1982.452 & 0.33595 & 17.9 \\
 & 5 & 1995.181 & 0.00003 & 30.59362 & 0.77630 & 1962.898 & 26.8 & 0.04361 & 0.21721 & -0.41962 & 1996.628 & 0.37549 & 15.9 \\
156026 & 3 & 1995.280 & 0.00002 & 48.45743 & 0.96648 & 1968.338 & 25.7 & -0.12203 & -176.96925 & 0.04332 & 2024.654 & 0.90965 & 25.6 \\
160346 & 5 & 1997.407 & 0.00028 & 12.73184 & 0.90492 & 1989.259 & 12.3 & -0.05695 & 1.20040 & -0.15781 & 1997.560 & 0.31172 & 11.0 \\
166620 & 0 & 1967.490 & 0.00014 & 12.75945 & 1.10908 & 1961.428 & 5.7 & 0.00851 & 1.79696 & -0.50082 & 1970.188 & 0.18084 & 5.6 \\
\multirow[t]{2}{*}{201092} & 3 & 1982.343 & 0.00103 & 11.44072 & 1.08425 & 1976.439 & 46.1 & 0.09411 & -0.22063 & 1.13625 & 1985.064 & 0.95599 & 46.1 \\
 & 4 & 1987.593 & 0.00044 & 16.39994 & 0.77715 & 1973.197 & 38.4 & 0.58815 & 4.87670 & 0.07907 & 1990.070 & 0.44106 & 37.7 \\
 \hline \hline
\end{tabular}%
}
\end{center}
\tablecomments{Cycle \#0 refers to the initial partial cycle. 
}
\end{table*}

\begin{table*}[]
\caption{Peak S-index, rise times, and rates of all usable cycles in each star.}
\begin{center}
\label{tab:WE_table}
\begin{tabular}{lccccccc}
\hline \hline
\multirow[t]{2}{*}{HD} & \multirow[t]{2}{*}{Cycle \# }& \multicolumn{3}{c}{quasi-Planck fit} & \multicolumn{3}{c}{skewed-Gaussian fit} \\
\cline{3-8}
 &  & Peak & rRate & rTime & Peak & rRate & rTime \\
\hline
\multirow[t]{2}{*}{3651} & 1 & 0.18337 & 0.00193 & 4.08333 & 0.19471 & 0.01116 & 1.91667 \\
 & 2 & 0.17585 & 0.00247 & 3.83333 & 0.17905 & 0.00026 & 2.08333 \\
\multirow[t]{3}{*}{4628} & 1 & 0.24400 & 0.01588 & 2.25000 & 0.24917 & 0.01348 & 2.08333 \\
 & 2 & 0.26172 & 0.02163 & 1.66667 & 0.27203 & 0.03091 & 1.50000 \\
 & 3 & 0.24875 & 0.01791 & 2.41667 & 0.25958 & 0.01308 & 1.83333 \\
\multirow[t]{3}{*}{10476} & 1 & 0.21654 & 0.01350 & 2.33333 & 0.22465 & 0.01241 & 2.16667 \\
 & 2 & 0.20730 & 0.01069 & 2.08333 & 0.21786 & 0.01813 & 1.41667 \\
 & 3 & 0.19456 & 0.00826 & 2.33333 & 0.19759 & 0.00531 & 2.00000 \\
\multirow[t]{2}{*}{16160} & 1 & 0.24944 & 0.01710 & 2.25000 & 0.25428 & 0.00967 & 3.58333 \\
 & 2 & 0.25672 & 0.01113 & 3.58333 & 0.25907 & 0.00648 & 2.75000 \\
\multirow[t]{3}{*}{20630} & 1 & 0.36284 & 0.02250 & 1.25000 & 0.37473 & 0.04399 & 0.75000 \\
 & 2 & 0.35601 & 0.00769 & 0.75000 & 0.36101 & 0.01294 & 1.58333 \\
 & 3 & 0.36306 & 0.01197 & 1.16667 & 0.37314 & 0.02452 & 1.83333 \\
\multirow[t]{3}{*}{26965} & 1 & 0.21072 & 0.00774 & 1.83333 & 0.21165 & 0.00897 & 1.75000 \\
 & 2 & 0.22545 & 0.01620 & 1.91667 & 0.22616 & 0.01746 & 1.91667 \\
 & 3 & 0.21582 & 0.01123 & 2.50000 & 0.21701 & 0.00920 & 2.83333 \\
\multirow[t]{2}{*}{32147} & 1 & 0.30300 & 0.01330 & 1.41667 & 0.31186 & 0.02212 & 1.25000 \\
 & 2 & 0.29910 & 0.01324 & 4.50000 & 0.30410 & 0.00876 & 2.58333 \\
\multirow[t]{4}{*}{81809} & 1 & 0.17986 & 0.00888 & 1.75000 & 0.18012 & 0.01053 & 1.50000 \\
 & 2 & 0.18189 & 0.00650 & 1.75000 & 0.18483 & 0.01131 & 1.66667 \\
 & 3 & 0.18231 & 0.00704 & 1.33333 & 0.18307 & 0.00833 & 1.41667 \\
 & 4 & 0.18238 & 0.00550 & 2.66667 & 0.18698 & 0.00310 & 3.91667 \\
\multirow[t]{4}{*}{103095} & 1 & 0.20525 & 0.01074 & 1.50000 & 0.20838 & 0.01698 & 1.75000 \\
 & 2 & 0.19617 & 0.00879 & 1.75000 & 0.19825 & 0.00922 & 2.25000 \\
 & 3 & 0.19601 & 0.00886 & 1.50000 & 0.19750 & 0.01035 & 1.08333 \\
 & 4 & 0.20207 & 0.01112 & 1.75000 & 0.20594 & 0.00705 & 2.91667 \\
\multirow[t]{2}{*}{115404} & 1 & 0.56161 & 0.00977 & 0.66667 & 0.57110 & 0.02425 & 1.50000 \\
 & 2 & 0.52006 & 0.02029 & 1.91667 & 0.52050 & 0.02214 & 1.83333 \\
\multirow[t]{7}{*}{149661} & 1 & 0.40591 & 0.05149 & 1.25000 & 0.41952 & 0.05803 & 1.83333 \\
 & 2 & 0.35677 & 0.02584 & 0.33333 & 0.35710 & 0.02737 & 0.33333 \\
 & 3 & 0.34502 & 0.01548 & 1.16667 & 0.34380 & 0.01250 & 1.25000 \\
 & 4 & 0.33207 & 0.01449 & 1.00000 & 0.33428 & 0.03913 & 0.66667 \\
 & 5 & 0.37405 & 0.02100 & 1.25000 & 0.37266 & 0.01355 & 0.91667 \\
 & 6 & 0.34182 & 0.01714 & 1.75000 & 0.36786 & 0.00247 & 3.50000 \\
 & 7 & 0.39966 & 0.03977 & 1.16667 & 0.39984 & 0.03958 & 1.16667 \\
\multirow[t]{3}{*}{152391} & 1 & 0.41087 & 0.01995 & 3.00000 & 0.42532 & 0.00200 & 3.91667 \\
 & 2 & 0.41707 & 0.02461 & 2.50000 & 0.41659 & 0.02560 & 1.66667 \\
 & 3 & 0.40970 & 0.01370 & 1.25000 & 0.43589 & 0.05929 & 1.00000 \\
\multirow[t]{4}{*}{155885} & 1 & 0.41012 & 0.00824 & 0.33333 & 0.41760 & 0.02513 & 1.66667 \\
 & 2 & 0.41478 & 0.01633 & 1.08333 & 0.41556 & 0.01378 & 1.75000 \\
 & 3 & 0.41979 & 0.00201 & 1.75000 & 0.44428 & 0.05823 & 0.66667 \\
 & 6 & 0.42966 & 0.03844 & 1.75000 & 0.48630 & 0.00001 & 2.45833 \\
\multirow[t]{2}{*}{155886} & 1 & 0.43646 & 0.02399 & 1.58333 & 0.46509 & 0.01321 & 3.08333 \\
 & 4 & 0.43613 & 0.04431 & 1.91667 & 0.43289 & 0.04327 & 1.58333 \\
\multirow[t]{2}{*}{156026} & 1 & 0.91094 & 0.02916 & 6.00000 & 0.91910 & 0.01062 & 7.25000 \\
 & 2 & 0.74860 & 0.02655 & 2.50000 & 0.74832 & 0.02086 & 2.50000 \\
\multirow[t]{4}{*}{160346} & 1 & 0.36610 & 0.04774 & 1.58333 & 0.36466 & 0.05019 & 1.41667 \\
 & 2 & 0.31719 & 0.01900 & 1.41667 & 0.32746 & 0.01845 & 2.16667 \\
 & 3 & 0.30662 & 0.01231 & 0.91667 & 0.32181 & 0.01240 & 0.83333 \\
 & 4 & 0.34045 & 0.03788 & 1.58333 & 0.36669 & 0.07513 & 1.91667 \\
\multirow[t]{5}{*}{161239} & 1 & 0.14510 & 0.00129 & 4.33333 & 0.14706 & 0.00050 & 6.08333 \\
 & 2 & 0.14098 & 0.00261 & 1.08333 & 0.14304 & 0.00599 & 1.08333 \\
 & 3 & 0.14238 & 0.00565 & 1.16667 & 0.14449 & 0.00582 & 0.41667 \\
 & 4 & 0.13780 & 0.00182 & 1.00000 & 0.13795 & 0.00201 & 1.00000 \\
 & 5 & 0.13834 & 0.00517 & 1.25000 & 0.14251 & 0.00642 & 0.58333 \\
\multirow[t]{2}{*}{166620} & 1 & 0.21458 & 0.00561 & 1.83333 & 0.22238 & 0.01183 & 2.91667 \\
 & 2 & 0.19361 & 0.00663 & 2.58333 & 0.20417 & 0.00312 & 3.66667 \\
\multirow{4}{*}{201091} & 1 & 0.78728 & 0.07623 & 2.00000 & 0.82616 & 0.10021 & 2.41667 \\
 & 2 & 0.69158 & 0.03955 & 1.41667 & 0.67393 & 0.03894 & 2.16667 \\
 & 3 & 0.70488 & 0.07721 & 1.50000 & 0.70868 & 0.09043 & 1.33333 \\
 & 4 & 0.75438 & 0.06348 & 2.08333 & 0.75525 & 0.06386 & 2.25000 \\
\multirow[t]{3}{*}{201092} & 1 & 1.05982 & 0.03324 & 0.75000 & 1.06719 & 0.02885 & 0.75000 \\
 & 2 & 1.13756 & 0.08544 & 3.25000 & 1.12020 & 0.01596 & 6.75000 \\
 & 5 & 1.06192 & 0.11032 & 1.33333 & 1.10408 & 0.29693 & 0.41667 \\
\multirow[t]{4}{*}{219834B} & 1 & 0.22465 & 0.00771 & 0.91667 & 0.23147 & 0.01977 & 1.83333 \\
 & 2 & 0.22145 & 0.00908 & 2.50000 & 0.22766 & 0.00852 & 3.58333 \\
 & 3 & 0.21575 & 0.01109 & 1.66667 & 0.22971 & 0.02723 & 1.08333 \\
 & 4 & 0.25346 & 0.03018 & 1.75000 & 0.25908 & 0.04676 & 2.16667 \\
 \hline \hline
\end{tabular}
\end{center}
\end{table*}


\end{document}